\documentclass[a4paper,11pt]{article}
\pdfoutput=1
\usepackage{jheppub}
\RequirePackage{luatex85}

\usepackage{amsmath, amssymb, mathrsfs}

\usepackage{graphicx}
\usepackage{dcolumn}
\usepackage{bm}
\usepackage{color}
\usepackage{cleveref}

\title{Foreground cleaning and template-free stochastic background extraction for LISA}

\date{\today}

\author[a]{Mauro Pieroni\,,}
\author[b,c]{Enrico Barausse\,}

\affiliation[a]{Theoretical Physics, Blackett Laboratory, Imperial College, London, SW7 2AZ,
United Kingdom}
\affiliation[b]{SISSA, Via Bonomea 265, 34136 Trieste, Italy and INFN Sezione di Trieste}
\affiliation[c]{IFPU - Institute for Fundamental Physics of the Universe, Via Beirut 2, 34014 Trieste, Italy}

\abstract{Based on the rate of resolved stellar origin black hole and neutron star mergers measured by LIGO and Virgo, it
is expected that these detectors will also observe an unresolved Stochastic Gravitational Wave Background (SGWB) by the time
they reach design sensitivity. A background from the same class of sources also exists in the LISA
band, which will be observable by LISA with 
signal-to-noise ratio (SNR) $\sim 53$. Unlike the stochastic signal from Galactic white dwarf binaries,
for which a partial subtraction is expected to be possible by exploiting its yearly modulation (induced by the motion of the LISA
constellation), the background from unresolved stellar origin black hole and neutron star binaries
acts as a foreground for other stochastic signals of cosmological or astrophysical origin, which may also be present in the LISA band.
Here, we employ a principal component analysis to model and extract an additional hypothetical  SGWB in the LISA band, without making any a priori assumptions on its spectral shape. At the same time,
we account for the presence of the foreground from stellar origin black holes and neutron stars, as well as
for possible uncertainties in the LISA noise calibration.
We find that our technique leads to a linear problem and  is therefore suitable for fast and reliable extraction of SGWBs 
with SNR up to ten times weaker than the foreground from black holes and neutron stars, quite independently of
the SGWB  spectral shape.}

\begin{document}
	
\maketitle
\flushbottom

\section{Introduction}
The Laser Interferometer Space Antenna (LISA)~\cite{Audley:2017drz} is a space-based Gravitational Wave (GW) detector scheduled to be launched by ESA, with junior
partnership from NASA, around 2034. LISA is currently in Phase A (assessment of feasibility), and among the items that are being investigated is the
scientific return as function of the mission design. The LISA observatory will be comprised by three spacecraft trailing the Earth around the Sun (by 10--15 degrees),
on an equilateral configuration with arms of about 2.5 million km. Along these arms, laser beams will be exchanged to monitor changes in the proper distance 
between the free-falling test masses carried by the spacecraft, in order to detect GW signals with frequencies in the mHz band.

The GW sky at such low frequencies is expected to be much more crowded than in the band $\gtrsim 10$--$100$ Hz accessible
from the ground~\cite{TheLIGOScientific:2014jea, Advanced-Virgo, Hild:2010id}. Indeed, LISA sources are intrinsically long-lived, i.e. they are expected to be active for the whole duration
of the mission (which is nominally of 4 yr), and they are foreseen to  number in the tens of thousands~\cite{Audley:2017drz} (including resolved sources alone).
In more detail, the strongest sources, with Signal to Noise Ratios (SNRs) up to thousands, are expected to be the mergers of massive
black hole binaries (with masses $\sim 10^4$--$10^7 M_\odot$),  which could be between a few and hundreds depending on the (currently unknown) astrophysical formation scenario~\cite{Sesana2004,Sesana2005,Barausse2012,Klein2016,2019MNRAS.486.4044B}. Extreme mass ratio   inspirals comprised of a massive black hole (with mass again $\sim 10^4$--$10^7 M_\odot$) and a smaller satellite stellar-origin
black hole (with mass $\sim 10$--$50M_\odot$) or neutron star may also be present, with very uncertain rates ranging from a few up to hundreds per year~\cite{Babak2017}.

LISA is also guaranteed to be able to observe a handful of white dwarf binaries that have already been observed in the electromagnetic band. These are referred to
as ``verification binaries''~\cite{verification}, and are just the tip of the iceberg of a much broader population of Galactic and extragalactic white dwarf binaries that are potentially
observable with LISA. Indeed, models of this population in the Milky Way predict that LISA should detect the signal of tens of thousands of resolved
Galactic white dwarf binaries at low frequencies~\cite{WD1,WD2,WD3,WD4,WD5}. Moreover, even more such sources will not have sufficient SNR to be detectable individually, but
will nevertheless constitute a formidable stochastic background signal, which  will be potentially strong enough to degrade the mission's 
sensitivity at low frequencies, i.e. to act as a foreground for other sources~\cite{cornish1,Adams:2013qma}. 

After the first LIGO detection of GW150914~\cite{Abbott:2016blz}, it
was realized that binaries of stellar-origin black holes, which merge in the band of existing or future ground based interferometers, will
also be detectable in LISA much earlier (by a time ranging from weeks to years) than their merger~\cite{Sesana:2017vsj} (see also \cite{Wong:2018uwb, Gerosa:2019dbe}). While the presence of
these sources is a blessing, since their long low frequency inspiral allows for measuring
their parameters to high accuracy, thus  permitting  precise tests of their formation scenario~\cite{Nishizawa1,Nishizawa2,Tamanini:2019usx}, of General Relativity~\cite{dipole,Carson:2019rda,Gnocchi:2019jzp}, and possibly identify
their electromagnetic counterparts (if any)~\cite{caputo_sberna}, many of them may not be detectable individually. The unresolved signal
from stellar-origin black hole binaries is indeed foreseen to be present in the LIGO/Virgo band  too, and should be detected 
when those detectors reach design sensitivity. Based on the current estimates of this 
background~\cite{LIGOScientific:2019vic}, its SNR in the LISA band is expected to be
around 53\footnote{This value is obtained (via~\cref{eq:SNR_definition}) by extrapolating the current estimates of the background
from stellar origin black hole and neutron star binaries in the LIGO/Virgo band to the LISA band.}
While this is not enough to degrade
the detector's sensitivity to resolved sources (unlike the case of Galactic white dwarf binaries), this
unresolved signal is expected to be certainly detectable once all resolved sources have been subtracted, by looking at the 
auto-correlation of the data residuals.

Unlike the stochastic background from Galactic binaries, however, the background from stellar-origin black holes is 
expected to be overwhelmingly extra-galactic in origin. As a result, while the unresolved signal from  Galactic binaries
can for the most part be subtracted by exploiting its yearly modulation (induced by the constellation's motion around the Sun)
and its anisotropy~\cite{Adams:2013qma},
the background from stellar-origin black holes will be (to a very good approximation) isotropic and stationary. This makes its subtraction difficult,
and may hamper the detection of other, weaker SGWBs that may be present in the data.

Indeed, one of the most important goals of the LISA mission is the possible detection of SGWBs of exotic origin. While
some of these, if present, may be very strong (e.g. stochastic signals from the superradiance driven spin down of the astrophysical black hole population, which may occur in models of fuzzy dark matter~\cite{Brito1,Brito2} or in the presence of exotic physics near the event horizon~\cite{Barausse:2018vdb}), most signals
of cosmological origin are expected to be rather weak. Indeed, it is possible to construct scenarios in which phase transitions in the early Universe~\cite{Caprini:2019egz}, networks of topological defects (e.g. cosmic strings)~\cite{Auclair:2019wcv} and even inflationary constructions~\cite{Bartolo:2016ami} may induce a sufficiently large SGWB to reach the sensitivity of LISA (for a review of these models see for example~\cite{Caprini:2018mtu} and reference therein). However, for these relatively weak signals, the background from stellar-origin black holes will act as
a foreground and may jeopardize their detection, unless it is successfully subtracted.

Here,  we introduce a template free approach for the detection and reconstruction of the LISA SGWB. Our method is based on a Principal Component Analysis (PCA; or singular value decomposition), and we show that it allows for the simultaneous detection and characterization of the foreground from stellar-origin black holes and  a signal of unknown origin and spectral shape. While similar in spirit to~\cite{Karnesis:2019mph, Caprini:2019pxz}, where template free approaches were also put forward, our technique has  the advantage of being very fast, because the determination of the posteriors becomes equivalent to solving a linear problem, and allows for 
detecting signals up to ten times weaker than the one from stellar-origin black holes. 

This paper is organized as follows: in~\cref{likelihood_sec}  we describe the procedure to generate our data and the technique on which our analysis hinges. In~\cref{sec:results} we show the results obtained by applying these techniques to a set of mock signals. Finally, in~\cref{sec:conclusions} we draw our conclusions. This work contains two appendices: in~\cref{sec:noise_fg} we describe the noise model used in our analyses and in~\cref{sec:appendixdownsampling} we discuss the procedure to downsample our data.

\section{A principal component analysis for the stochastic background}\label{likelihood_sec} 

We start by making the standard simplifying assumption that once all resolved sources have been subtracted, the LISA 
data (i.e. the residuals) in a given channel are described by  a stationary Gaussian process~\footnote{Residual non-Gaussianities 
and non-stationarities may be present due to instrumental glitches and the subtraction procedure of thee resolved sources~\cite{Ginat:2019aed}.
To account for this, it is
possible to include additional parameters in the model for the power spectral density.}.
If we deconvolve the LISA data with the response function of the detector, as described in \cref{sec:noise_fg},
the resulting data $d(t)$ will satisfy
\begin{equation}\label{psd}
\langle d(f) d^*(f')\rangle=S(f)\delta(f-f')\,,
\end{equation}
where \begin{equation}d(f)=\int_{-\infty}^{\infty}  d(t) \exp(-2 \pi i f t)dt\end{equation}
 is the (complex) Fourier transform  of the time domain data $d(t)$,
 $\langle\ldots\rangle$ denotes
an ensemble average (i.e. an average over many data realizations), and $S(f)$ is referred to as the (double sided) 
 spectral density of the data $d$~\cite{Maggiore:1900zz}. 

Since LISA samples the data for a finite
observation time $T$, the Fourier transform is defined only at discrete frequencies $f_i$ (with $i$ the frequency index ranging e.g. from $1$ to $n$)
spaced by $\Delta f=1/T$, and \cref{psd} becomes
\begin{equation}\label{psd2}
\langle d_i d^*_j\rangle=\frac{1}{\Delta f} S(f_i)\delta_{ij}\,,
\end{equation} 
where $d_i\equiv d(f_i)$.
Since the residuals are Gaussian, the real and imaginary parts of the $d_i$ obey a Gaussian distribution
with variance set by \cref{psd2}, i.e.
\begin{equation}
\label{eq:gaussian}
p({\rm Re}\,d_i,{\rm Im}\,d_i)= \frac{1}{\pi S_i}e^{-[({\rm Re}\,d_i)^2+({\rm Im}\,d_i)^2]/S_i}\,.
\end{equation}
By changing variables to the absolute value ($|d_i|$) and phase 
of the data, one obtains that the phase is uniformly distributed, while $|d_i|$ is described 
by $p(|d_i|)=2|d_i| e^{- |d_i|^2/S_i}/S_i$.

Since different frequencies are uncorrelated, as per \cref{psd}, the likelihood can be obtained
by multiplying the probability distributions of the various sampled frequencies $f_i$. In particular,
by working not with $|d_i|$ but with $D_i\equiv |d_i|^2$, one obtains
\begin{equation}\label{likelihood0}
p(D_{i=1,...,n}|S)= \prod_i^n\frac{1}{S_i}e^{-D_i/S_i}\,,
\end{equation}
for the data in each channel.
At a fixed frequency $f_i$, the mean value of $D_i$ is $\mu_i=S_i$ and its variance
$\sigma^2_{i}=\mu_i^2= S_i^2$.

We now consider $\bar{D}_i$, which we define as the average of $D_i$ (with $i$ fixed) over $N\gg 1$ chunks in which we divide the time series.
By the central limit theorem, we can approximate the probability distribution 
function for $\bar{D}_i$ with a Gaussian centered in $\mu_i=S_i$ and with variance $\sigma^2_{i}/N=\mu_i^2/N=S_i^2/N$.
The likelihood for the averaged data $\bar{D}_i$ then becomes
\begin{equation}\label{likelihood}
p(\bar{D}_{i=1,...,n}|S)\approx \frac{N^{n/2}}{(2\pi)^{n/2}}\prod_{i=1}^n  \frac{1}{S_i}e^{-\frac{N(\bar{D}_i-S_i)^2}{2S_i^2}}\,.
\end{equation}
We can simplify the likelihood (which eventually will allow us to make the problem linear) by further noting that $\bar{D}_i\approx S_i$ near the peak of the likelihood, which
allows for writing
\begin{equation}\label{likelihood2}
p(\bar{D}_{i=1,...,n}|S)\approx \frac{N^{n/2}}{(2\pi)^{n/2}}\prod_{i=1}^n  \frac{1}{\bar{D}_i}e^{-\frac{N(\bar{D}_i-S_i)^2}{2\bar{D}_i^2}}\,.
\end{equation}

Let us assume that the power spectral density $S$ is the sum of three contributions, from the signal, instrumental noise
and astrophysical foregrounds, i.e.
\begin{equation}\label{tot}
S(f)=S_{\rm signal}(f)+S_{\rm noise}(f)+S_{\rm foreground}(f)\,.
\end{equation}
For the signal, let us now assume the 
form 
\begin{equation}
\label{eq:signal_model}
S_{\rm signal}(f)=\sum_{j=1}^{m} a_j \delta_w(f-f^a_j)\,, 
\end{equation}
where the $a_j$, $j=1,...,m$ are parameters to be determined, $f^a_j$ are pivot frequencies associated with the $a_j$, and the functions $\delta_w(F)$ are defined by
\begin{equation}\label{deltaw}
\delta_w(F)=\frac{1}{\sqrt{2\pi} w} e^{-\frac12 F^2/w^2} \,.
\end{equation}
Clearly, by evaluating \cref{eq:signal_model} at the sampled frequencies $f_i$, one obtains
\begin{equation}\label{model}
S^{\rm signal}_i=S_{\rm signal}(f_i)=\sum_{j=1}^{m} a_j \delta_w(f_i-f^a_j)\,. 
\end{equation}

 Note that in the limit $w\rightarrow 0$, one has $\delta_w(F) \rightarrow \delta(F)$. Therefore, by choosing $w\rightarrow 0$, 
$f^a_j = f_i$ and  $m = n$, the parameters $a_j$ would simply be the reconstructed values of the signal at the sampled frequencies. 
Nevertheless, a non-zero $w$ allows for encoding the fact that the signal is expected to be a smooth function of $f$.
Moreover, allowing for $f^a_j \neq f_i$ and $m$ potentially lower than $n$ allows for extra flexibility in the
technique and will prove useful in practice, as we show in Sec.~\ref{sec:results}. Indeed,
we will show that  in some situations it is useful to
work with a ``reduced'' basis, i.e. set $m\ll n$, as this
turns out to enhance stability of the results and also decreases
the dimensionality of the problem.
Indeed, one may even consider the number of
coefficients to be used as a free parameter of the model, and estimate its optimal
value within a Bayesian framework (i.e. by analyzing its posterior distribution or by using other criteria, as we discuss in the following).
While  we  leave this possibility for future work, in Sec.~\ref{sec:results} we briefly
investigate how our results change when the dimension of the working basis is reduced.

In practice, instead of working with the parameters
$a_j$, we utilize the rescaled parameters $\alpha_j=a_j/K_j$, $j=1,...,m$,
where the arbitrary normalizations $K_j$ are chosen to make the parameters $\alpha_j$ dimensionless and (as much as possible) of order
unity.\footnote{A possible choice is to take $K_j\sim | \bar{D}(f^a_j) - S_{\rm noise,exp}(f^a_j) -S_{\rm foreground,exp}(f^a_j) |$,
with $\bar{D}(f^a_j)$ set to the data $\bar{D}_i$ with frequency $f_i$ closest  to $f^a_j$,
and with $S_{\rm noise,\,exp}$ and $S_{\rm foreground,\,exp}$ the 
values of the instrumental noise and LIGO/Virgo foregrounds expected based on (i.e. maximizing) the priors. These
are given by \cref{noise} and \cref{fg} below, with $A=O=L=1$. We stress, however, that other prescriptions for the $K_j$ are also possible and our results are robust with respect
to the specific choice made.}  We find that this improves the numerical stability of the technique.
Clearly, this is obviously equivalent to multiplying ${\delta_w(f-f_j)}$ by $K_j$. We then assume uniform priors for the parameters $\alpha_j$.

For the instrumental noise, let us assume that we know the functional form
of the acceleration and optical metrology system contributions~\cite{Caprini:2019pxz}, save for two normalization coefficients $A$ and $O$~\footnote{The $A$ and $O$ parameters used in this work are respectively the square of $A$ and of $P/10$ of~\cite{Caprini:2019pxz}. Note that more complicated/different uncertainties in the spectral shape and amplitude of
the instrumental noise can also be incorporated in our  technique (although care needs to be used to keep the problem linear
if computational cost is an issue). However, irrespective of the adopted extraction technique, the 
detection of any SGWB with LISA crucially relies on informative priors on the instrumental noise, because
cross correlation techniques such as those  used in LIGO and Virgo are not applicable to LISA
(since noise in the three arms is correlated).},
for which we assume Gaussian priors of about 20\% width~\cite{LISA_docs}:
\begin{align}\label{noise}
&S_{\rm noise}(f)=A \, S_{\rm acc}(f)+O \, S_{\rm OMS}(f)\,,\\
&p(A,O)\propto e^{-\frac12 [(A-1)^2/\sigma_A^2+(O-1)^2/\sigma_O^2]}\,,
\end{align}
with  $\sigma_A=\sigma_O=0.2$, while the functional forms $S_{\rm acc}(f)$ and  $S_{\rm OMS}(f)$ are given in \cref{sec:noise_fg}.

For the astrophysical foreground, we assume that the spectral shape of
the signal from binaries of stellar origin black holes and neutron stars is known (see \cref{sec:noise_fg} for details) up to a normalization coefficient $L$. We can thus write
\begin{align}\label{fg}
&S_{\rm foreground}(f)=  L \, S_{\rm LV}(f) \,,\\
\label{eq:prior_fg}
&p(L)\propto e^{-\frac12 [ (L-1)^2/\sigma_L^2]}\,,
\end{align}
where we assume $\sigma_L=0.5$. Present~\cite{ligostoch, Martynov:2016fzi} and upcoming~\cite{Punturo:2010zz, Sathyaprakash:2012jk, Maggiore:2019uih} ground based detectors are actually expected to measure this foreground, providing more stringent constraints on $L$. However, in order to test the robustness of our method, we use a relatively large value for $\sigma_L$. Furthermore, as an additional test for robustness, we checked that less stringent choices for the prior leave the results shown in~\cref{sec:results} unaffected.

 As mentioned previously, we neglect here the contribution of the
foreground from Galactic binaries, since its time dependence (caused by the motion
of the detector) and it anisotropy allow for measuring its power spectral density to high precision (and thus
for removing it)~\cite{Adams:2013qma}. In any case, a residual contribution of the
population of  Galactic binaries could be accounted for simply by adding a suitable term to eq.~\eqref{fg},
and does not significantly affect the results.

With these assumptions, the posterior probability distribution
for the parameters $\boldsymbol{\theta}~=~(\{\alpha_j\}_{j=1,\ldots,n}, A, O, L)$
is given by Bayes' theorem and reads
\begin{align}
&\ln p(\alpha_{j=1,...,n_{\rm max}},A,O,L|\bar{D}_{i=1,...,n})= -\frac{\chi^2}{2} +{\rm const}\,,\\
&\chi^2=\sum_{i=1}^n  {N\frac{(\bar{D}_i-S_i)^2}{\bar{D}_i^2}} +\frac{(A-1)^2}{\sigma_A^2}+\frac{(O-1)^2}{\sigma_O^2}
+\frac{(L-1)^2}{\sigma_L^2}
\,,\label{chi2}\\
&S_i=\sum_{j=1}^{n} \alpha_j K_j \delta_w(f_i-f^a_j)+A S_{\rm acc}(f_i)+O S_{\rm OMS}(f_i) +  L S_{\rm LV}(f_i)
\end{align}
Since $S_i$ depends linearly on the parameters,
finding the maximum of the posterior distribution (or equivalently the minimum $\chi^2$) is a linear problem, i.e. one has to solve the linear system $\partial \chi^2/\partial \theta_j=0$ with $j=1,...,m+3$.
Note that for  $m+3>n$, the problem becomes degenerate, since there are more
parameters than data. However, 
even for $m+3\leq n$ 
the problem may still be degenerate in practice, especially for large $m$, since we do not expect to be able to extract all the parameters reliably due to the errors affecting the data.  
This issue can be bypassed by performing a PCA of the Fisher matrix $F_{ij}$, as we will outline
in the following.

The Fisher matrix is defined by
\begin{equation}
F_{ij}= \frac12\frac{\partial^2 \chi^2}{\partial \theta_i \partial \theta_j}\,,
\end{equation}
with $i,j=1,\ldots,m+3$. This matrix, which  is independent of the parameters since the problem is linear, encodes their errors and correlations.
Solving for the eigenvalues and eigenvectors of the Fisher matrix then allows for 
identifying linear combinations of the parameters that are uncorrelated with one another,
as well as computing their errors. In more detail, one can
define the functions $\{\eta_{(i)}(f)\}_{i=1,...,m+3}$ with
\begin{align}\label{basis}
\eta_{(i)}(f) & =\eta^{\rm signal}_{(i)}(f)+\eta^{\rm noise}_{(i)}(f) +\eta^{\rm LV}_{(i)}(f)\,\\
\eta^{\rm signal}_{(i)}(f) & =\sum_{j=1}^{m} e^{(i)}_j K_j\delta_w(f-f_j^a)\,\\
\eta^{\rm noise}_{(i)}(f) & =e^{(i)}_{m+1}S_{\rm acc}(f)+e^{(i)}_{m+2}S_{\rm OMS}(f)\,\\
\eta^{\rm LV}_{(i)}(f) & =e^{(i)}_{m+3}S_{\rm LV}(f)
\end{align}
where the vectors $\{\vec{e}_{(i)}\}_{i=1,...,m+3}$ are  $n$ \textit{orthonormal} eigenvectors of the Fisher matrix. Note that if $m+3 > n $, at least $m-n+3$ of these eigenvectors will correspond to vanishing eigenvalues, since the problem is degenerate. 
However, even if $m+3 \leq n $ many eigenvalues may be very small. This is central for the PCA/singular value decomposition technique, 
as will become clear below.

The total power spectral density given 
by eq.~\eqref{tot} can then be re-written as a sum on the functions $\{\eta_{(i)}(f)\}_{i=1,...,m+3}$, i.e.
\begin{align}\label{model2}
S_i&=S(f_i)=\sum_{k=1}^{m+3} b_k \eta_{(k)}(f_i)\,,\\
b_k&=\sum_{j=1}^m \alpha_j e^{(k)}_j+ A e^{(k)}_{m+1}+ O e^{(k)}_{m+2}
+Le^{(k)}_{m+3}\,.
\end{align}
The coefficients $b_k$ are now uncorrelated Gaussian variables, and their errors are given by 
$\lambda_{(k)}^{-1/2}$, where $\lambda_{(k)}$ is the eigenvalue corresponding to the eigenvector $\vec{e}_{(k)}$. In particular, by using eq.~\eqref{basis} we can write the reconstructed signal, noise spectral density and astrophysical foregrounds as
\begin{align}\label{model3}
S^{\rm signal}_i&=\sum_{k=1}^{m+3} b_k \eta^{\rm signal}_{(k)}(f_i)\,,\\
S^{\rm noise}_i&=\sum_{k=1}^{m+3} b_k \eta^{\rm noise}_{(k)}(f_i)\,,\label{model3bis}\\
S^{\rm LV}_i&=\sum_{k=1}^{m+3} b_k \eta^{\rm LV}_{(k)}(f_i)\,.\label{model3tris}
\end{align}
As already mentioned, we can then perform 
a PCA to ``de-noise'' the reconstructed quantities, 
i.e. one can rewrite eqs.~\eqref{model3}--\eqref{model3tris} by including only
the coefficients $b_k$ which are ``well determined'' (\emph{e.g.} one possibility is to only include coefficients 
whose values are not compatible with zero at one $\sigma$, namely $|b_k| > \lambda_{(k)}^{-1/2}$). This yields
\begin{align}\label{model4}
S^{\rm signal}_i &=\sum_{|b_k| > \lambda_{(k)}^{-1/2}} b_k \eta^{\rm signal}_{(k)}(f_i)\,,\\
S^{\rm noise}_i &=\sum_{|b_k| > \lambda_{(k)}^{-1/2}} b_k \eta^{\rm noise}_{(k)}(f_i)\,,\label{model4bis}\\\
S^{\rm LV}_i &=\sum_{|b_k| > \lambda_{(k)}^{-1/2}} b_k \eta^{\rm LV}_{(k)}(f_i)\,.\label{model4tris}
\end{align}
Since the $b_k$ are uncorrelated Gaussian variables, the (Gaussian) errors on $S^{\rm signal}_i$, $S^{\rm noise}_i$ and $S^{\rm LV}_i$ can then be obtained
by summing in quadrature the errors of the $b_k$ (i.e. $\lambda_{(k)}^{-1/2}$), with coefficients given by these equations. These are
the errors  for the reconstructed signal shown in Figs.~\ref{fig:large_flat}--\ref{fig:bump} below. 

Finding explicitly the eigenvalues and eigenvectors of the Fisher matrix can be challenging, since
the matrix is singular or almost singular, and the dimensionality of the parameter space
is huge. Indeed, $m$ can be as large as the number of sampled frequencies $n$, which
for LISA could be of the order of $10^6$, because the data sampling rate $\Delta t$ will be of the order of a few seconds.\footnote{Note that even though the nominal duration of the LISA mission
will be 4 yr, our technique assumes that the data are divided in $N\gg 1$  chunks, in order  to write the likelihood of \cref{likelihood}
and \cref{likelihood2}.} The first problem can be addressed in practice by 
replacing $F_{ij}\to F_{ij}+\epsilon \delta_{ij}$. This make
the Fisher matrix formally non-singular, with eigenvalues $\gtrsim \epsilon$. As long
as $\epsilon$ is much smaller than the minimum $b_k$ contributing to the sum in eqs.~\eqref{model4}--\eqref{model4tris}, the reconstructed ``cleaned'' model is unaffected.
As for the high dimensionality of the problem, the computational cost may 
be reduced by noting that the Fisher matrix becomes sparse in a suitable basis. In more  detail,
one can write
\begin{align}
F_{ij}=& \sum_{k=1}^m\frac{N}{\bar{D}_k^2}\frac{\partial S_k}{\partial \theta_i} \frac{\partial S_k}{\partial \theta_j}+ \frac{1}{\sigma_A^2}\delta_{i}^{m+1}\delta_{j}^{m+1}
+\frac{1}{\sigma_O^2}\delta_{i}^{m+2}\delta_{j}^{m+2}
+\frac{1}{\sigma_L^2}\delta_{i}^{m+3}\delta_{j}^{m+3}
\,,
\end{align}
where 
\begin{align}
\frac{\partial S_k}{\partial \theta_i}&=\sum_{l=1}^m K_l\delta_w(f_k-f_l^a) \delta^{l}_{i}+S_{\rm acc}(f_k)\delta^{m+1}_i +S_{\rm pos}(f_k)\delta^{m+2}_i
+S_{\rm LV}(f_k) \delta^{m+3}_i \, .
\end{align}
From this expression, it is clear that unless  $w\Delta t\gg 1$, the Fisher matrix becomes sparse -- with non-zero elements only near the diagonal or in the
last four rows/columns  -- which might allow for decreasing the burden of the computation
when real data are available.

Nevertheless, for the purpose of this work, where we are just concerned with applying
our technique to mock simulated data, we downsample the latter in the following manner.
We start by bundling the $n$ data $\bar{D}_i$ in groups of $M$ 
data adjacent in frequency. Since the variance of $\bar{D}_i$
is $S_i^2/N$ [c.f.~eq.~\eqref{likelihood}], it seems natural to replace each group by the weighted average
 $\bar{\bar{D}}_k=(\sum_{i=1+kM}^{(k+1)M} \bar{D}_i/S_i^2)/(\sum_{i=1+kM}^{(k+1)M} 1/S_i^2)$, 
and to assign this value to the frequency
$\bar{f}_k \equiv (\sum_{i=1+kM}^{(k+1)M} f_i/S_i^2)/(\sum_{i=1+kM}^{(k+1)M} 1/S_i^2)$.
Indeed, in~\cref{sec:appendixdownsampling} we show that the
best estimate of the (total) power spectral density $\bar{S}_k= S(\bar{f}_k)$ at the
frequency $\bar{f}_k$ is indeed given by the  ``downsampled'' data $\bar{\bar{D}}_k$,
with a (Gaussian) variance approximately given by $\bar{S}_k^2/M\approx\bar{\bar{D}}_k/M$.  
Therefore, eqs.~\eqref{likelihood}--\eqref{likelihood2} remain valid
for the ```downsampled'' data, modulo the replacement $N\to NM$, as one would intuitively expect. The rest of the analysis proceeds unchanged, again
with $N\to NM$. Note that to compute the weighted averages needed to 
define the $\bar{\bar{D}}_k$ and the $\bar{f}_k$, one can replace $S_i\approx \bar{D}_i$,
as we did when going from 
eq.~\eqref{likelihood} to eq.~\eqref{likelihood2}. 

For all the analyses presented in the next section, we consider frequencies in the range $f \in [10^{-4} , 2 \times 10^{-2} ]\text{Hz}$, which in logarithmic scale corresponds to a region which is roughly symmetric around to the peak of the LISA sensitivity (which is around $2 \times 10^{-3} $Hz). We start with by generating $N=94$ Gaussian realizations\footnote{We assume that the data are divided into chunks of around 11 days each (with a total observation time of 4 years and $75\%$ duty cycle). Note that the number of chunks controls how well the central limit theorem 
holds when going from \cref{likelihood0} to \cref{likelihood}. Since $N$ is large but finite, we expect some small bias to be 
potentially present in our results. Further bias may be introduced when approximating \cref{likelihood} with \cref{likelihood2}, which allows for 
making the problem  linear, at the expense of neglecting the skewness  of \cref{likelihood} (which disappears in
\cref{likelihood2}). We have verified that all these biases are reduced by choosing larger $N$.} 
of $d(f)$ obeying~\cref{psd2} with an initial spacing of $10^{-6}$Hz, and we then set $M =10$ to downsample our data.

\section{Results}
\label{sec:results}
We can now proceed to demonstrate the robustness and effectiveness of the procedure described in the previous section. To this purpose, we apply our technique to a set of mock signals with various SNRs. Note that depending on the particular shape of the input signal, a different ``correlation length'' $w$, as
defined in \cref{deltaw}, will be required for capturing the features of the signal. 
For the analysis shown in this section, we choose $w$ by ``trial and error'',
i.e. by choosing the value best suited for recovering the injected signal. In reality,
when the injected signal is unknown, different correlation lengths should be tested in order to determine the one best describing the shape of the unknown SGWB. This corresponds to finding the $w$ giving the largest posterior probability (\emph{i.e.} the smallest $\chi^2$), which
can be found simply by varying $w$ on a uniform grid ``by brute force''. A similar procedure can
be applied to ascertain the optimal dimensionality $m$ of the basis, as defined in the previous section.
That would entail varying $w$ and $m$ on a grid, and then applying our technique to each point, in order 
to find which one gives the lowest $\chi^2$ (c.f. sec. 4.2.1 of \cite{DAbook} for a problem solved by a similar technique). 
Alternatively, Markov Chain
Monte Carlo techniques can be used to sample the posterior distribution. Another possibility  for model comparison, which can also keep track of the number of the degrees of freedom employed in the analysis, is the Akaike Information Criterion (AIC)~\cite{Akaike}.
These analyses are however beyond the scope of this work, in which we instead choose $w$ and $m$ by trial and error.

\begin{figure}[h!]
	\includegraphics[width=0.99\textwidth]{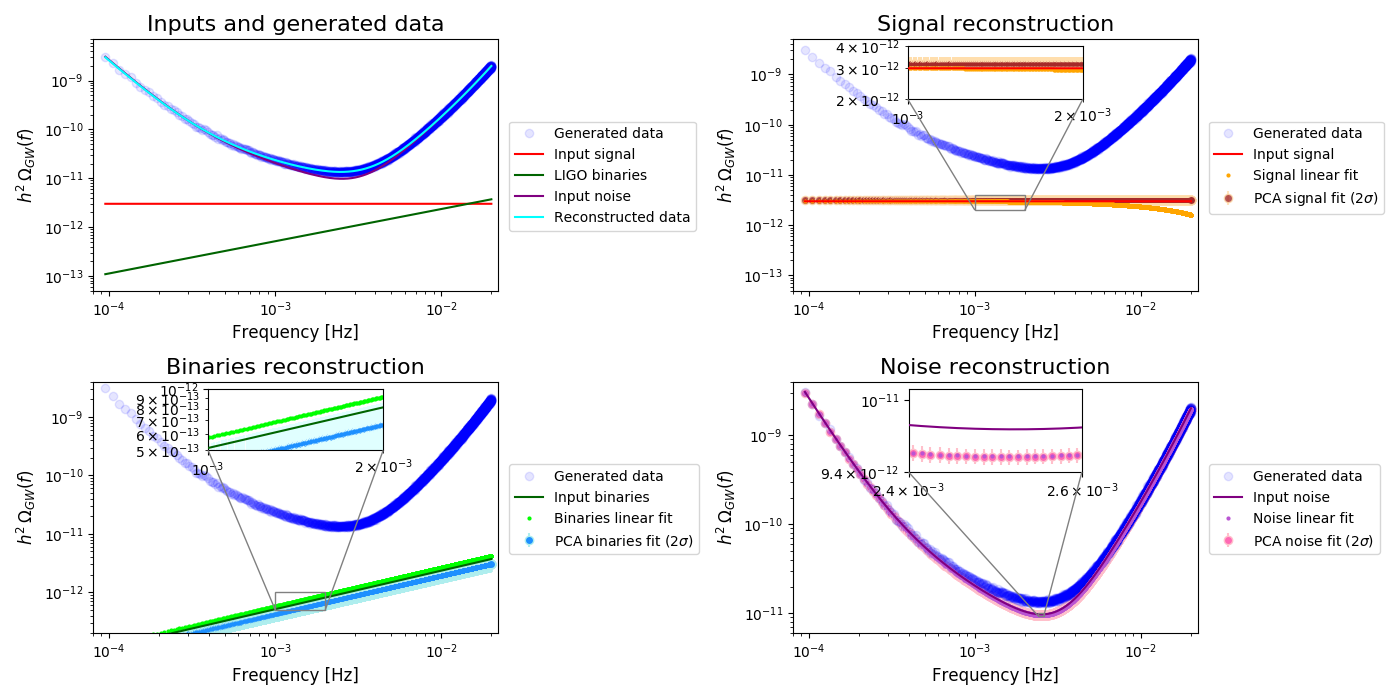}
	\caption{PCA analysis for a flat input signal with amplitude $h^2 \Omega_{*}=3\times 10^{-12}$ and SNR $\simeq 156$. For this analysis we have fixed $m=10$ and $w = 1$Hz (see main text for details). \emph{Top left}: a plot of all inputs, the simulated data and the reconstructed data. \emph{Top right}: input signal, linear fit and PCA reconstructions with $2\sigma$ error bands. \emph{Bottom left}: input LIGO/Virgo foreground (see main text for details), linear fit and PCA reconstructions with $2\sigma$ error bands. \emph{Bottom right}: input LISA noise (see main text for details), linear fit and PCA reconstructions with $2\sigma$ error bands. 	\label{fig:large_flat}}
\end{figure}

As a first example, we consider in~\cref{fig:large_flat} a flat background with amplitude $h^2 \Omega_{*}=3 \times 10^{-12}$, corresponding to SNR $\simeq 156$. Note that in this case the injected signal is larger than the foreground from LIGO/Virgo binaries in most of the frequency range. Moreover, since this signal is constant in the LISA band,  it is appropriate to choose a large correlation length $w$, but that causes a large degeneracy between $S_{LV}$ and $S_{\rm signal}$. This makes it
numerically difficult to disentangle the two components. An efficient way to resolve this issue is to  reduce the dimensionality $m$ of the signal basis\footnote{Clearly, a lower $m$ must be compensated by a larger $w$ in order for the basis to remain sufficiently dense to accurately model the signal.} In the analysis presented in~\cref{fig:large_flat}, we choose a basis of $m = 10$ Gaussians (with a uniform logarithmic spacing for the pivot frequencies $f^a_i$) and  $w=1$Hz. 
As can be seen from~\cref{fig:large_flat}, which shows the  PCA reconstructed signal, foreground and noise, this choice
very efficiently disentangles the different components hidden in the data.
The reconstructed value for the LIGO/Virgo foreground parameter is $L \simeq 1.044 \pm 0.114$. As for the LISA noise parameters, we obtain $A \simeq 0.980 \pm 0.005$ and $O \simeq 0.976 \pm 0.001$.\footnote{Note  that the parameters $A$ and $O$ are significantly different from 1. As mentioned earlier, this bias comes about
because the central  limit theorem, used to go from \cref{likelihood0} to \cref{likelihood}, only holds 
in the limit in which the number of chunks $N$ diverges. We have indeed verified that by increasing $N$, $A$ and $O$ 
become compatible with $1$. The same applies to the cases shown in the figures below.}  Concerning the reconstruction of the signal, we can see from the top right panel that the linear fit
 (defined as the model minimizing the $\chi^2$ in \cref{chi2}, or equivalently by  the sum given in eqs.~\eqref{model3}--\eqref{model3tris}, where all coefficient are included, irrespectively of their value)
 correctly reconstructs the signal in most of the frequency range. However, at large frequencies, where the foreground is larger than the signal, it fails. On the other hand the PCA reconstruction, by dropping low information components, produces a fairly accurate reconstruction of the signal even in this range.

\begin{figure}[h!]
	 \includegraphics[width=0.99\textwidth]{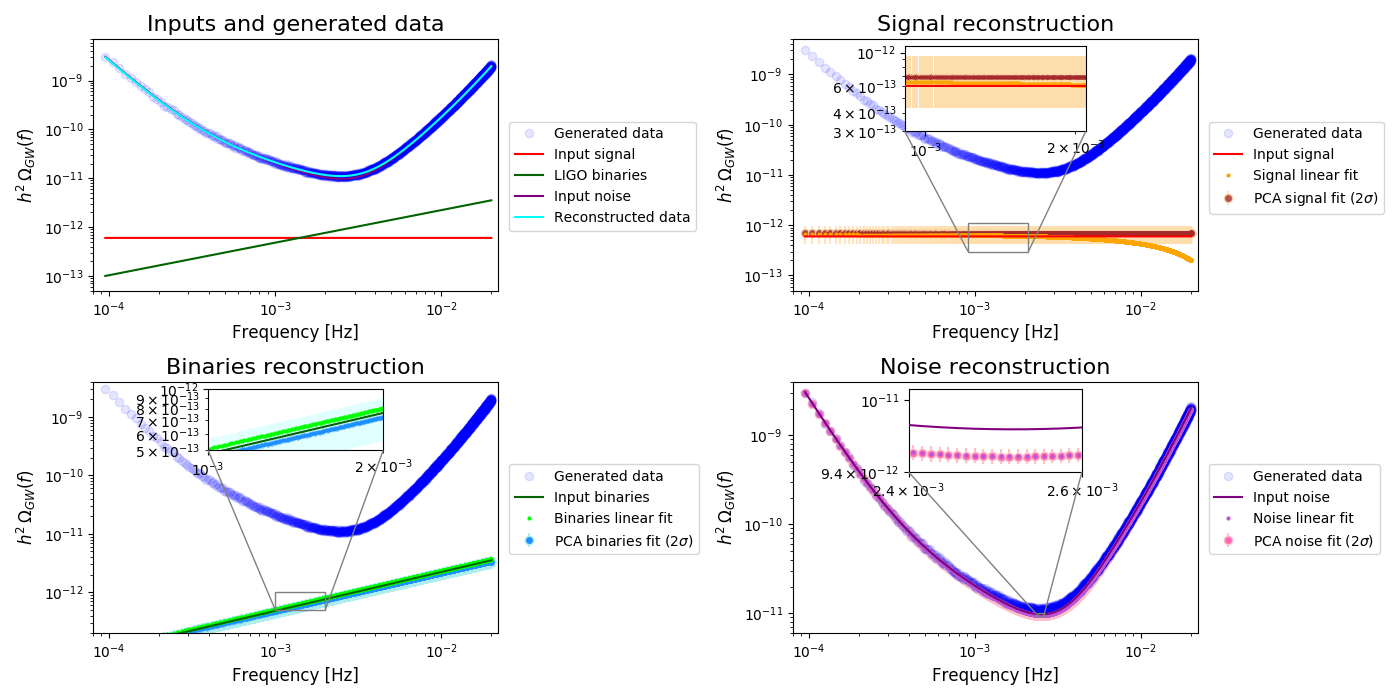}
	\caption{PCA analysis for a flat input signal with amplitude $h^2 \Omega_{*}=6 \times 10^{-13}$ and SNR $\simeq 31$.  This analysis was performed with $m=10$ and $w = 1$Hz. Plot structure as in~\cref{fig:large_flat}. \label{fig:low_flat}}
\end{figure}

In~\cref{fig:low_flat}, we consider the opposite situation, \emph{i.e.} we consider an injected signal  smaller than the LIGO/Virgo foreground in most of the frequency range. In more detail, we choose a flat signal with amplitude $h^2 \Omega_{*}=6\times10^{-13}$, corresponding to an SNR $\simeq 31$. Consistently with the analysis of~\cref{fig:large_flat}, we have used  $w = 1$Hz and $m= 10$ (with the same uniform logarithmic spacing for the pivot frequencies). We can see that while in this case the reconstruction the LISA noise parameters is basically unaffected ($A \simeq  0.978 \pm 0.005$ and $O \simeq 0.975 \pm 0.001$), 
the determination of LIGO/Virgo foreground parameter is slightly more accurate ($L \simeq 0.896 \pm 0.106$). As for the signal reconstruction, the top right panel of~\cref{fig:low_flat} clearly shows that, despite the signal being much smaller than the LIGO/Virgo foreground, our procedure still captures it in most of the frequency range. Once again the PCA reconstruction works better than a simple linear fit in the region where the signal is much weaker than the foreground.

\begin{figure}[h!]
	\includegraphics[width=0.99\textwidth]{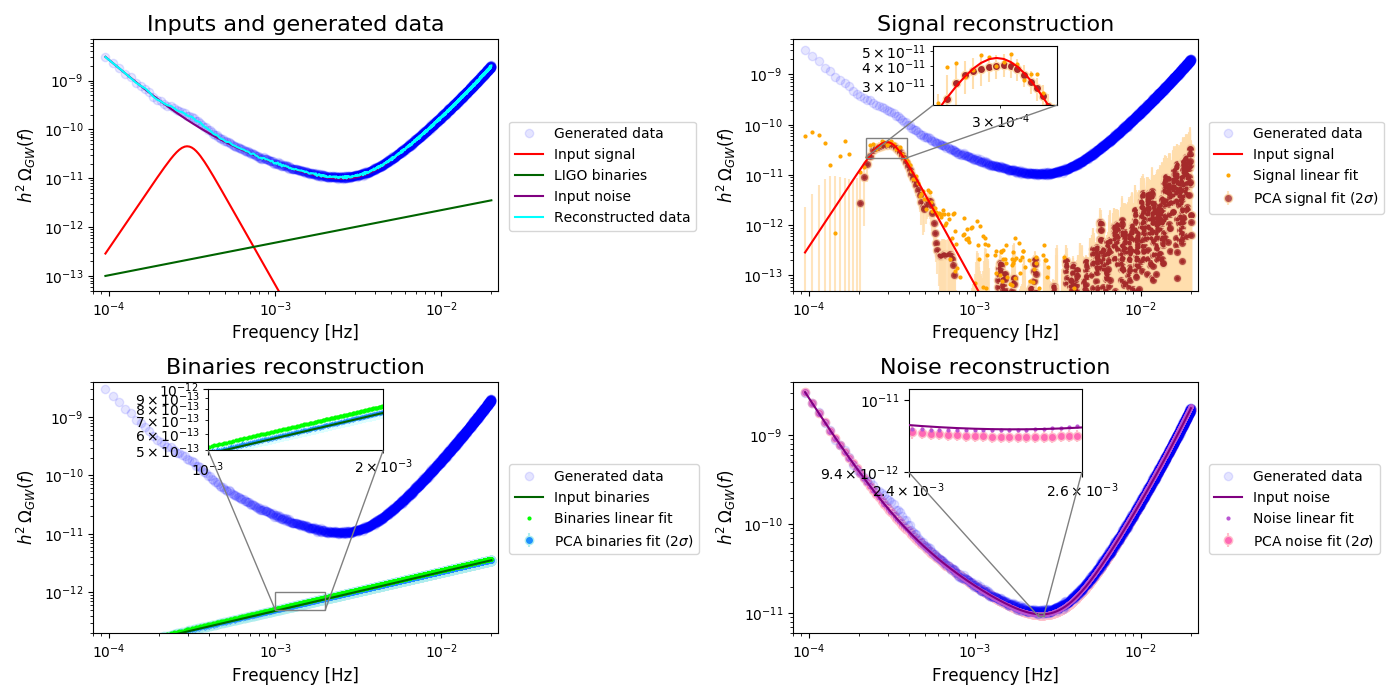}
	\caption{PCA analysis for an input broken power law signal (see~\cref{footnote:broken}) with amplitude $h^2 \Omega_{*}=9 \times 10^{-11}$, $n_{s1} = 5$, $n_{s2} = -6$ and $f_* = 3 \times 10^{-4}$Hz. The SNR of this signal  is $\simeq 34$. This analysis was performed with $m=n$, $w=2\times10^{-5}$Hz. Plot structure as in~\cref{fig:large_flat}. 	\label{fig:broken}}
\end{figure}

\begin{figure}[h!]
	\includegraphics[width=0.99\textwidth]{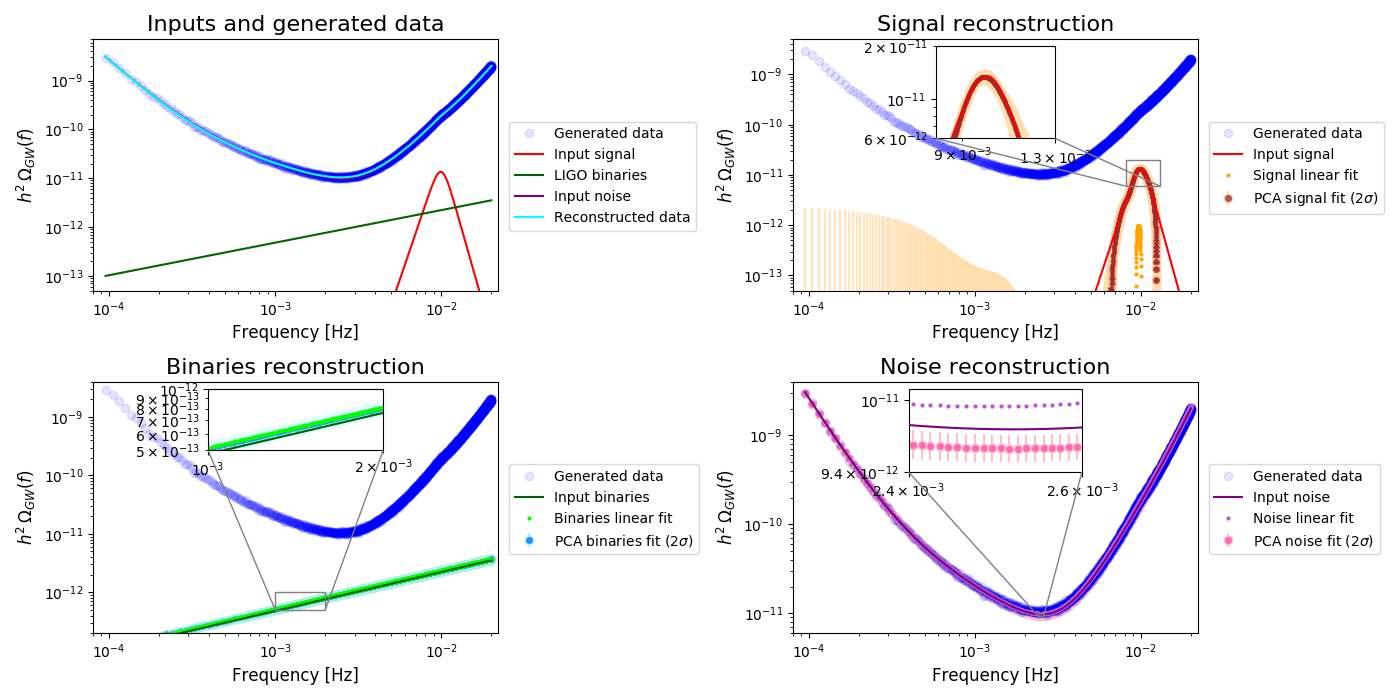}
	\caption{PCA analysis for an input broken power law signal (see~\cref{footnote:broken}) with amplitude $h^2 \Omega_{*}=9 \times 2.7 \times 10^{-11}$, $n_{s1} = 10$, $n_{s2} = -12$ and $f_* = 10^{-2}$Hz.  The SNR of this signal  is $\simeq 34$. This analysis was performed with $m=100$ with uniform log spacing and $w = 5 \times 10^{-4}$Hz. Plot structure as in~\cref{fig:large_flat}.
		\label{fig:bump} }
\end{figure}

We conclude this section by showing two cases where the input signal is chosen to be less degenerate with the foreground. In~\cref{fig:broken} we show the results obtained by applying our procedure ($m=n$, $w=2\times10^{-5}$Hz) to a broken power law signal~\footnote{\label{footnote:broken} The signal is given by
  $h^2 \Omega_{GW} (f) =h^2 \Omega_*  \, ({f}/{f_*})^{n_{s1}} /[1 +({f}/{f_*})^{n_{s1} -n_{s2}} ]$, where $ \Omega_* $ is the amplitude, $f_*$ is the pivot frequency, and $n_{s1}$, $n_{s2}$ are respectively the spectral tilts before and after the pivot. } placed at small frequencies
and with SNR $\simeq 34$. In~\cref{fig:bump} we show instead the results obtained with $m=100$ with uniform log spacing and $w=5 \times 10^{-4}$Hz on an injected 
signal given by 
 a broken power law placed at high frequencies and with  SNR $\simeq 34$. In both cases, the reconstruction of the LIGO/Virgo foreground is accurate ($L \simeq 0.948 \pm 0.034$ and $L \simeq 0.989 \pm 0.045$ respectively), as is the case for the LISA noise parameters ($A \simeq 1.000 \pm 0.004$, $A \simeq 0.983 \pm 0.009$ and $O \simeq 0.981 \pm 0.001$, $O \simeq 0.984 \pm 0.002$ respectively).

\section{Discussion}
\label{sec:conclusions}

In this work we have proposed a template-free PCA technique for the reconstruction of SGWBs with  LISA. We have shown that our procedure is very efficient at disentangling an unknown input signal from the instrumental noise even in the presence of a foreground. Remarkably, the component separation obtained with the techniques presented in this work is effective even for signals with fairly small  SNR $\simeq30$, compared to a power-law foreground from
LIGO/Virgo binaries with SNR$\simeq 53$.

Our approach essentially consists of expanding the signal onto a basis of Gaussians with fixed width $w$ and centered
on a set of pivot frequencies within the LISA frequency band. Generically, an unknown signal can thus be expressed as a linear combination of these functions through some unknown parameters to be determined. We model 
uncertainties in the LISA noise budget and in the amplitude of 
the LIGO/Virgo  foreground  with global normalization parameters also to be determined. The best fit for all 
these parameters can be then be found by maximizing the posterior distribution given by Bayes' theorem. 
By dividing the time series of the data in chunks and using the central limit theorem,
we manage to reduce this  maximization procedure to a linear problem, which only requires
the inversion of the Fisher matrix of the parameters. This inversion
can be challenging because of the large dimensionality of the problem, which makes the matrix
singular or almost singular (i.e. most combinations of the parameters are essentially unconstrained by the data).

The PCA technique  provides in fact a robust technique to tackle this issue, by essentially dropping
all the linear combinations of the parameters on which the data provide no information, and retaining
only the ``components'' that are informed by the data. This procedure leads to a ``de-noised''
agnostic reconstruction of the underlying signal, as  well as of the LISA instrumental noise and astrophysical LIGO/Virgo foreground.

In conclusion, the method described in this work is efficient at disentangling the different components
of the SGWB that LISA will measure. Our results are of major relevance since they clearly show that foreground subtraction can be successfully performed. This would allow for the identification (and eventually for the characterization) of the possible cosmological signals that might be hidden behind the foreground. Moreover, since our technique is based on a template-free approach, it provides a robust tool applicable to any background signal. Indeed, while in this work we have restricted our analysis to LISA, our techniques can be easily extended to different detectors.

\begin{acknowledgments}
This paper is dedicated to the memory of  Pierre Bin\'etruy in  the third  anniversary of  his death. Among many more things, Pierre pushed
us both to think about stochastic backgrounds for LISA, and this work would not have seen the light of day without his example and inspiration. We thank Carlo Contaldi, Vincent Desjacques, Valerie Domcke, Nikolaos Karnesis, Antoine Petiteau and Angelo Ricciardone for useful comments and discussions. We acknowledge financial support provided under the European Union's H2020 ERC Consolidator Grant ``GRavity from Astrophysical to Microscopic Scales'' grant agreement no. GRAMS-815673. M.P. acknowledges the support of the Science and Technology Facilities Council consolidated grants ST/P000762/1. 
\end{acknowledgments}

\appendix

\section{The noise and foreground default models}
\label{sec:noise_fg}
In each of the  time-delay interferometry (TDI) channels $X$, $Y$ and $Z$\footnote{In this work, we focus on a single channel of the LISA data, for illustration purposes. However, the analysis can be readily extended to more than one channel.}, the LISA experiment will provide data $x(f)$ whose auto-correlation
\begin{equation}
\label{eq:data_strain}
\langle x(f) x^*(f')\rangle =\frac12 \delta(f-f') \left[\mathcal{R} (f)S_{\rm sign+fg}(f) +   P_{\rm noise}(f)\right] 
\end{equation} 
where  $\mathcal{R}(f)$ is the detector polarization- and sky-averaged 
response function, $S_{\rm sign+fg}(f)$ is the strain power spectral density of the signals and foregrounds, 
while $P_{\rm noise}(f)=P_{\rm acc}(f)+P_{\rm OMS}(f)$ is 
expected to be comprised of two contributions, one from the single mass acceleration noise and one from
the optical metrology system noise. In more detail, these two noise contributions are expected to be given by~\cite{LISA_docs}
\begin{equation}
\begin{aligned}
P_{\rm acc}(f) &=  16 \sin^2\left(\frac{2 \pi f L}{c}\right)   \left[ 3 +  \cos^2\left(\frac{4 \pi f L}{c}\right) \right]  \left(3\frac{{\rm fm}}{{\rm s}^2\,\sqrt{\rm Hz}}\right)^2 \times \\
& \qquad \times  \left[1 + \left(\frac{0.4\,\textrm{mHz}}{f} \right)^2  \right] \left[1 + \left(\frac{f}{8\,\textrm{mHz}} \right)^4  \right] \left(\frac{1}{2 \pi f } \right)^4 \left(\frac{2 \pi f}{c} \right)^2 \; , \\
P_{\rm OMS}(f) &=  16 \sin^2\left(\frac{2 \pi f L}{c}\right)   \left(15\frac{{\rm pm}}{\sqrt{\rm Hz}}\right)^2\left[1 + \left(\frac{2\,\textrm{mHz}}{f} \right)^4  \right] \left(\frac{2 \pi f}{c} \right)^2\; , 
\label{eq:noise-AB}
\end{aligned}
\end{equation}
while the response function  can be well approximated by~\cite{Caprini:2019pxz,LISA_docs}
\begin{equation}\label{eq:resp}
\mathcal{R}(f) \simeq 16 \sin^2\left(\frac{2 \pi f L}{c}\right)\frac{3}{10} \frac{1}{1+0.6(2\pi f L /c)^2} \left(\frac{2 \pi f L}{c}\right)^2\,
\end{equation}
where $L=2.5$ Gm is the arm length.
Note that this simplified noise budget relies on the assumption that
the constellation's arm lengths are equal and constant, and that noise contributions of the same kind
present the same power spectral density. As explained in the main text, we account for this uncertainties by
introducing two parameters $A$ and $O$ rescaling the amplitudes of the acceleration and optical metrology power spectral densities, 
with Gaussian priors of about 20\% width.

In this paper, rather than working directly with the data $x(f)$, we work with $d(f)\equiv \sqrt{2}x(f)/\sqrt{\mathcal{R} (f)}$, in terms of which
\cref{eq:data_strain} becomes
\begin{equation}
\label{eq:data_strain2}
\langle d(f) d^{*}(f')\rangle = \delta(f-f') [S_{\rm sign+fg}(f) +  S_{\rm acc}(f)+S_{\rm OMS}(f)]\,,
\end{equation}
where we have defined
\begin{align}
S_{\rm acc}(f)=\frac{P_{\rm acc}(f)}{\mathcal{R} (f)}\,, \qquad \qquad
S_{\rm OMS}(f)=\frac{P_{\rm OMS}(f)}{\mathcal{R} (f)}\,.
\end{align}

As in the main text, we can then decompose $S_{\rm sign+fg}(f)$ into signal(s) and foreground(s). For the latter,
the contribution from Galactic binaries presents a yearly modulation that is expected to allow for its subtraction, and
thus focus only on the background from LIGO/Virgo binaries. At low frequencies, their signal is very well approximated 
by~\cite{Regimbau:2011rp,Maggiore:1900zz,LIGOScientific:2019vic}
\begin{align}
S_{\rm LV}(f)&=\frac{3 H_0^2}{4\pi^2 f^3} \Omega_{\rm GW}(f)\,,\\
\Omega_{\rm GW}(f) &= \Omega_{*} \left( \frac{ f}{f_{*}}\right)^{2/3}\,,
\end{align} 
where $H_0=h\, 100 {\rm km/(s Mpc)}$, $h= 0.679$~\cite{Aghanim:2018eyx}, and $\Omega_{*}$ -- i.e. the amplitude at the pivot frequency $f_*=25$ Hz -- will be measured by LIGO/Virgo at design sensitivity~\cite{ligostoch, Martynov:2016fzi}.\footnote{Although detailed studies are still missing, future ground based third generation detectors~\cite{Punturo:2010zz, Sathyaprakash:2012jk, Maggiore:2019uih}, which may be online at the same time or slightly after LISA,
will also provide cogent information on $S_{\rm LV}$. Indeed, they may measure most binaries contributing to $S_{\rm LV}$ as resolved sources,
besides measuring it in part as an unresolved component.}
As our default value, we assume here $\Omega_{*}=8.9\times 10^{-10}$, which is the current best estimate 
based on the measured rate of coalescence of resolved neutron star and stellar origin black hole binaries in 
LIGO/Virgo~\cite{LIGOScientific:2019vic}.\footnote{As explained in \cite{LIGOScientific:2019vic}, this estimate of $\Omega_{*}$ assumes a Salpeter mass function for the primary black hole
in a binary, while the secondary is drawn from a uniform distribution. For neutron star binaries,
each component is drawn from
a Gaussian distribution with mean of $1.33 M_\odot$  and 
standard deviation of $0.09M_\odot$. The assumed intrinsic merger rates are
$56\,$Gpc$^{-3}\,$yr$^{-1}$ and $920\,$Gpc$^{-3}\,$yr$^{-1}$, respectively for black holes and neutron stars, and correspond
to the best estimates from the GstLAL pipeline.}
As mentioned in the main text, to show the robustness of our methods 
we use a conservative estimate of $\sigma_L = 0.5$ for the width of our prior in~\cref{eq:prior_fg}. 

Finally, we recall that the SNR  of a SGWB can be computed as
\begin{equation}
{\rm SNR} \equiv \sqrt{T \int_{f_{\text{min}}}^{f_{\text{max}}} d f \, \left( \frac{S_h (f)}{S_n(f)}  \right)^2 }\,, 
\label{eq:SNR_definition}
\end{equation}
where $f_{\text{min}}$, $f_{\text{max}}$ are respectively the minimum and maximum frequencies to which the detector is sensitive, $T$ is the observation time (which we set in this paper to 4 yr with $75\%$ efficiency), and $S_h$ 
and $S_n$  are respectively the SGWB and detector noise power spectral densities.

\section{Downsampling the data}
\label{sec:appendixdownsampling}
In this Appendix, we provide additional details on our downsampling procedure.
As mentioned in the main text, let us start by bundling the data in groups of $M$
points adjacent in frequency. If the frequency range spanned by these $M$ data is sufficiently 
small, we can locally approximate the power spectral density within this range by performing 
a linear fit. To this purpose, it is convenient to introduce the reference frequency
$\bar{f} \equiv (\sum_i f_i/S_i^2)/(\sum_i 1/S_i^2)$ (where the index $i$ spans the $M$ points
under consideration), express the data in terms of the distance from this
reference frequency, i.e. $\delta f= f-\bar{f}$, and fit the data with $S=a\delta f+b$.

By assuming uniform priors on $a$ and $b$, and using the fact that the data are 
independent Gaussian variables if eq.~\eqref{likelihood} is valid, Bayes' theorem returns the same answer as the standard $\chi^2$ fit 
procedure, i.e. the best estimates for $a$ and $b$ are~\cite{DAbook}
\begin{gather}
a=\frac{\beta p-\gamma q}{\alpha\beta-\gamma^2}\,,\quad
b=\frac{\alpha q-\gamma p}{\alpha\beta-\gamma^2}\\
\alpha=\sum_i \omega_i\delta f_i^2\,,\quad
\beta=\sum_i \omega_i\,,\quad \gamma=\sum_i \omega_i\delta f_i\,\quad p=\sum_i \omega_i\delta f_i \bar{D}_i\,,\quad q=\sum_i \omega_i \bar{D}_i\,,
\end{gather} 
where the weights $\omega_i=2/\sigma_i^2$ depend on the errors of the data,
$\sigma_i=S_i/\sqrt{N}$. Note that from the definition of the $\delta f_i$ as distances from the reference frequency
$\bar{f}$, it follows that $\gamma=0$. The best estimate for $b=S(f=\bar{f})\equiv \bar{S}$ is therefore $b={q}/{\beta}$,
which coincides with the weighted average of the data, $\bar{\bar{D}}$, defined
in the main text.

Moreover, the posterior distribution for the parameters $a$ and $b$ is a Gaussian with covariance matrix~\cite{DAbook}
\begin{align}
\sigma^2_{aa}&=\frac{2\beta}{\alpha\beta-\gamma^2}\,,\\
\sigma^2_{bb}&=\frac{2\alpha}{\alpha\beta-\gamma^2}\,,\\
\sigma^2_{ab}&=\sigma^2_{ba}=-\frac{2\gamma}{\alpha\beta-\gamma^2}\,.
\end{align}
Since $\gamma=0$ in our case, $b$ is uncorrelated from $a$ and its posterior distribution
is Gaussian with variance $\sigma^2_{bb}=2/\beta$. This means that the (Gaussian) variance
to be ascribed to $\bar{\bar{D}}$ is $(\sum_i 1/\sigma^2_i)^{-1}=N^{-1}(\sum_i 1/S^2_i)^{-1}$, which
we can approximate by $\bar{S}^2/(N M)$ if the $S_i$ do not vary strongly.

\bibliographystyle{JHEP}

\bibliography{shortbib}

\end{document}